# Understanding The Chemical Bond – A Simple View

Robert B. Murphy* and Richard P. Messmer*

**Abstract**
In keeping with, the perhaps apocryphal, dictum of Einstein: "If you can't explain it simply, you don't understand it well enough," we attempt a simple explanation of chemical bonding behavior based on the fundamental physical basis that underlies chemistry - namely Quantum Mechanics. We present the basics to non-experts in quantum theory who want to have some fundamental concepts that help them think about experiments and interpret them with reliable basic concepts. The focus is on hyper-valent molecules containing sulfur atoms and uses the simplest correlated wave function that allows the bonds to be localized between two atoms. This contrasts with the MO and DFT theories which yield orbitals that are delocalized over the molecule. We illustrate the proposed simple concepts with quantum calculations on several sulfur containing molecules and note the transferability of the bonds and bonding orbitals from molecule to molecule.

**Introduction**
Quantum Mechanical (1) calculations on several sulfur-containing molecules (2,3,4) using the simplest correlated description of local bonding between two atoms (5) are presented. The contrast between the proposed approach and that of Molecular Orbital (MO) (6) and Density Functional Theory (DFT) (7) are discussed. The proposed method allows predictions to be made about bond lengths and bond angles before calculations are carried out. Using Effective Core Potentials (ECP, 8), the computations can be carried out on a laptop computer. An important realization is that the Li-Ne row of the periodic table is unique because of the small size of its atomic cores. The volume of the atomic cores for the Na - Ar atoms is an order of magnitude larger than the volume of the cores for the Li - Ne atoms (8). This turns out to be critical in understanding the difference between S and O chemistry as well as the differences in bonding between the elements of the Li row and the Na row of the periodic table. This was pointed out also for P and N in a paper in 1990 as discussed in the conclusions.

We provide detailed calculations in the "Results" section in support of our thesis regarding "hyper-valency." But here, consider the following test calculations on the S and O atoms. We maintain that due to the small size of the atomic core for the Li-Ne row of the periodic table that it is not possible to have six orbitals about the O atom core (for example) in contrast to the situation for the S atom. Remember the core volume of S is an order of magnitude greater than for O (8). Our hypothesis regarding the importance of the core size is easily tested. Suppose we start with the S atom and try to put 6 valence orbitals about that atom. This can be done by having all the singly occupied orbitals orthogonal to one-another, as it would have to be in the $SF_6$ molecule, for example. The Pauli repulsion between the 6 orbitals around the O atom will be very large as compared to the S atom, because six orbitals will not fit around an O atom! In comparison the 6 orbitals about the S atom easily fit around the atom.

So, we calculate the total energies for the six orbitals about each of the S and O atoms and compare these energies. The computations show that the result for the O atom with 6 orbitals is 330 kcal/mol higher than for the S atom with 6 orbitals. The computational approach used for this comparison is the same as for the molecules we discuss later. The details of the approach are discussed in the section "Calculational Approach" in the Results.

In the following discussion it is convenient to introduce concepts and refine them in a series of "Rounds." Each Round is more accurate in its concepts but is based on the concepts from the previous Rounds. Many presentations of concepts mix concepts from several of our Rounds in starting their discussions. Hopefully the present approach, using Rounds will be easier to follow and understand.

It is important to stress who our target audience is for this paper, and who it is not targeted for. It is NOT targeted for experts in quantum mechanical calculations of electronic systems of interest in chemistry, physics, and materials science. For them, it may seem that our going back over the history of how the quantum mechanical concepts evolved is redundant. Their approaches are the standard ones which have come to be accepted as how to think about electronic structure and a belief that nothing new need be added. However, there is a large group of non-specialists who are looking for an intuitive understanding about electronic structure that might assist them in thinking about experimental results. They may find an alternative way of thinking useful in their goal of having a simple intuitive approach, in conjunction with the standard approaches – but which has a rigorous physical underpinning. It is for this latter group that this paper is intended. We believe this latter group may be much larger than the former.

## Round 1

To begin, let's start with the idea of "energy levels". One of the earliest ideas in Quantum Mechanics begins with the idea that there is a set of energy levels occupied by electrons that are arranged on a vertical energy scale. Consider an atom of Neon. It has 10 electrons and 5 energy levels. Why just 5 levels? Each energy level in simple quantum theory is occupied by 2 electrons. This is called the "Pauli exclusion principle" (10). No more than two electrons can occupy an energy level. And the two electrons are referred to as "spin-up" and "spin-down" electrons by established custom. Further the lowest level on the energy scale is referred to as the 1s level. The next level is referred to as the 2s level, followed by the 2p levels. The p levels always have 3 levels with the same energy (d levels consist of 5 levels with the same energy, etc.). This labelling is a matter of historical use and is not essential to our understanding - but is necessary to discuss with those who are fluent in the quantum mechanical language.

As we ascend the "energy levels" of an atom, molecule or solid, the spacing of energy levels from one level (or set of levels) to the next higher level(s) decreases. Hence the spacing between the 1s and 2s levels is greater than between the 2s and 2p levels.

There are also "unoccupied levels" in all atoms, molecules, and solids. These levels represent possible "excited states" of the system (be it atoms, molecules, or solids). For example, we can imagine an electron from the 2p orbitals of Neon moving into a 3s orbital to create an excited state of Neon. Or from the 2p orbitals into the 3p orbitals, which would be higher in energy.

By the construction of quantum theory, the orbitals in the stable ground state of the system all have negative energies on our energy scale. If an electron from the highest occupied level is excited to an energy level of zero energy, the energy difference approximates the "ionization energy."

As a system becomes larger in terms of the number of electrons it contains, the energy spacing between the highest energy levels of the system becomes much smaller. In solids the density of states requires a different approach in terms of the methods for calculating the properties of the system. This different approach is referred to as "band theory" (11).

The "simplest molecule" is the $H_2$ molecule. The molecule contains two electrons - one from each of the H atoms. Hence, we expect one occupied orbital for the molecule containing two electrons. This is the simplest case of a "chemical bond." So, the next question of chemical significance is how to describe the dissociation of the molecular chemical bond into hydrogen atoms. This is the essence of chemistry. How do we describe in simple terms the making and breaking of chemical bonds? The most common way of treating molecules by quantum theory is called the "molecular orbital" theory (6). It consists of a set of doubly occupied orbitals - just as in the case of the Neon atom discussed above.

There is an enormous amount of experimental data that bonds can be thought of as localized between two atoms (in the simplest cases). That is, a C-H bond is very much the same in $CH_4$ or $C_2H_6$, etc. or for a C-H bond in a protein. However, when molecular orbital (MO) theory is applied to molecules, the orbitals are delocalized over many of the atoms in the molecule, thus losing the notions inspired by a host of experimental findings.

Another issue is that in describing the dissociation of a chemical bond, i.e., moving two atoms or molecular fragments apart from each other, MO theory does not agree with the experimental observations regarding this fragmentation. The molecule does not dissociate into two fragments that agree with experimental observations. Although there are many approximate ways to alleviate this problem, they are not theoretically sound and do not lead to a consistent qualitative picture over many similar molecules.

For the $H_2$ molecule a rigorous theoretical approach was presented nearly one hundred years ago. We will describe this in the next Round. Unfortunately, this original approach for $H_2$ (12) resisted efforts to generalize it to arbitrary molecules for many decades. Many approximate methods developed, and the resultant concepts helped to formulate the "valence bond" (VB) approach to describing molecules. Such approaches required a lot more computational effort than was available to the non-theoretical specialists. Thus, the molecular orbital theory came to dominate the approaches to describing molecules for decades, although it lost the simple intuitive and experimentally supported notion of local chemical bonds. However, it was computationally viable and mathematically much simpler to understand.

## Round 2
The hydrogen atom is the only system that has a full quantum mechanical description. All other systems require approximations. The physical basis of the quantum mechanics necessary to understand chemistry is the "Schrödinger equation" (1) and, to repeat, is **only** solvable for the H atom. Hence a lot of the concepts used in understanding chemistry can be traced to generalizations and approximations based on the solution for the H atom.

The Schrödinger equation consists of two mathematic expressions, one for the "Hamiltonian" and the other for the "wave function." The Hamiltonian for H consists of two components - the "kinetic energy" and the "potential energy." There are small additional terms that are necessary to describe some physics not related to chemistry. The solution of the Schrödinger equation for the H atom, yields a ground state 1s wave function and a set of higher energy solutions, namely the 2s, 2p, 3s, 3p, 3d, etc.

Hence, we may think of the solution for the F atom (the F wave function) as a product of the 9 H orbitals: 1s1s' 2s2s' $2p_a 2p_a$' $2p_b 2p_b$' $2p_c$. The spacing in energy (potential energy) of the orbitals for the F atom will be determined from the orbital energies of the H atom. This follows the discussion of Round 1, but now we have a more rigorous understanding of how the orbital energies can be computed. Note that there is no interaction between the electrons (the Coulomb interaction for example) in this discussion of the F atom. The "total energy" of the F atom in this description is just the sum of the orbital energies occupied from the H atom orbital energies. However, there can be other approximations to determine the energy levels of a system (atoms, molecules, solids) more appropriate than using the H-atom orbitals and energies.

This system wave function (a product of one-electron orbitals based on the H atom) is a crude approximation to the "correct" wave function of a system from a rigorous perspective. To understand the rigorous wave function characteristics, it is useful to consider the hydrogen molecule. The electron is an "elementary particle" known as a "fermion." The rigorous fermion wave function for any system has a property that is called "antisymmetric."

If we consider the rigorous wave function for the hydrogen molecule, we have the following situation. Suppose we orient the molecule along the horizontal axis. Then we have a H nucleus to the left, denoted by *l*, and the electron on that atom is denoted by the label 1. Likewise, there is a H nucleus to the right, denoted by *r* and the electron on that atom is denoted by the label 2.

The accurate molecular electronic wave function is then denoted by equation 1:

$$N_s [\phi_l(1) \phi_r(2) + \phi_l(2) \phi_r(1)] [\alpha(1)\beta(2) - \beta(1)\alpha(2)] \tag{1}$$

where Ns is a normalization constant, $\phi$ are the H atomic orbitals or more generally - localized wave packets, and the $\alpha, \beta$ are electron spin functions. This is the equation as presented by Heitler and London in 1927 (12). It is referred to as the valence bond solution of the hydrogen molecule. This requires some explanation as it is the first example of the antisymmetric property of electrons.

The basic idea is that the electrons are indistinguishable from one another. So, if the electrons are exchanged between the two H atoms, there will be two terms included in the wave function. But these two terms in the wave function are required to combine in a fashion to satisfy the antisymmetric principle. For a system of N-electrons, irrespective of the number of atoms, there will be N! such terms. Where N! is "N factorial" – which is the product of N x (N-1) x (N-2) x (N-3) x … x 1); which is a very large number for a system of N electrons. Further, this means that each electron has its own orbital in contrast to Molecular Orbital (MO) theory where each orbital is occupied by two electrons. This exchange interaction is associated with an energy, the "exchange energy" (13), which is strictly a quantum effect with no classical analog. But it is the fundamental reason why the energy levels (orbitals) can only be occupied by one or two electrons, as proposed via the Pauli exclusion principle (10), see Round 1. Pauli originally proposed it, in an ad hoc way, to be consistent with experimental observations.

So, for comparison the MO theory spatial wave function for the $H_2$ molecule is given by:

$$[\varphi_l(1)+\varphi_r(1)] [\varphi_l(2)+\varphi_r(2)] = \varphi_l(1)\varphi_l(2)+ \varphi_r(1)\varphi_r(2)+ \varphi_l(1)\varphi_r(2)+ \varphi_r(1)\varphi_l(2). \qquad (2)$$

Note the first two terms in the expansion of this equation, where there are two electrons on each of the atoms. When the molecule dissociates (or any other set of two molecular fragments) these terms represent very "ionic contributions" to the energy which are not consistent with the molecule dissociating into known molecular fragments. These ionic contributions represent very high energy effects, inconsistent with experiment. Note also that the orbitals ϕ and φ in the two equations may or may not be the same.

So, MO theory is mainly used for describing systems around their experimentally determined geometries. There are some further approximations to MO theory that give partial ad-hoc remedies to the problem, however.

Another important approximation that is used in almost all calculations on electronic systems is the "Born – Oppenheimer approximation" (14). This approximation notes that the atomic nuclei are much heavier (by a factor of 1000 or more), than the electrons of the system and can be thought of as standing still while the electrons move about the atoms in the molecule. Hence, for each molecular geometry the nuclei are clamped in place and the wave function is determined for each geometry.

An important rigorous theorem which is critical to almost all calculations of electronic systems is the "variational principal." It states that for the ground state of the system – if the total energy of the system is considered as a function of the electronic wave function, then an approximation to the electronic wave function of the system can be improved if the system is lowered in energy by changing the guess of the approximation to the wave function. This principle is used in conjunction with the Born-Oppenheimer approximation. Methods for implementing the variational principle are both difficult and computationally demanding, having consumed a lot of theoretical effort over the last several decades. This has been aided by the rapid advances in computational capabilities. A key concept in such computations is the self-consistent field (15). The wave function for the system is constructed from the orbitals then a search is made, by changing the orbitals and hence the system wave function until the total energy is lowered. This procedure is repeated over and over until the total energy is no longer lowered. A consequence of the procedure for MO calculations is that the orbital energies, the spacing between them, and the orbitals change. These concepts will be dealt with in more detail in Round 3.

In summary, the basic points of Round 2, building on Round 1 with more rigor, are:
1) A lot of concepts related to chemical thinking are derived from the accurate solution of the Schrödinger equation for the hydrogen atom. The only system that has a rigorous solution to the Schrödinger equation.
2) The anti-symmetry of the system wave function has a very significant energy, called the exchange energy, which is the origin of the orbital energy spacings.
3) The MO solution for $H_2$, and all other electronic systems, has ionic terms that do not agree with experiment. The valence bond solution of $H_2$ is consistent with the antisymmetric principle and the proper dissociation into two H atoms. However, its generalization to chemical bonding in systems other than $H_2$ was not discovered for several decades.

## Round 3

In this Round, we start by considering the generalization of the Heitler-London treatment of the $H_2$ molecule to any covalent bond in any molecule. Although there are several Valence Bond approaches that have been developed, we will focus on a simple one originated by Goddard and co-workers (5), which is the strict generalization of the Heitler-London method. To get an intuitive feeling for what it is, we consider the methane molecule, $CH_4$. C has 4 valence electrons and 2 core electrons. The 4 valence electrons of C and the 4 electrons from the H atoms form 4 covalent C-H bonds. Each bond points to an apex of a tetrahedron. Let's consider two bonds and label them as (C1-H1) and (C2-H2). The most general Valence Bond description would also consider the C1-H2 and C2-H1 bonds and all the other possibilities of pairing the orbitals. These other spin pairings (or "resonance" structures) lower the energy of the system – but to a much, much smaller degree than the original pairing. This pairing of two electrons to form a covalent bond between two adjacent atoms (as in the $H_2$ molecule) is referred to as "perfect pairing." This "perfect pairing" (PP) for the bonds, seems like a reasonable approximation consistent with the notion of chemical bonds and requires much, much less computational effort. Now, in MO calculations, the orbitals are doubly occupied and "orthogonal" to one another. This greatly simplifies the calculations and doesn't seem to be a harmful approximation at equilibrium bond distances for closed shell molecules. Hence in the valence bond method we are discussing, this approximation of orbital orthogonality between valence bonded pairs is adopted and is referred to as strong orthogonality (SO). The valence bond (VB) method (5) we will focus on in the following discussions is referred to as the Generalized VB-SOPP method (GVB-SOPP). It takes account of static correlation effects only. A more general Valence Bond method (4), referred to as the Modern Valence Bond Theory, also accounts for some dynamic correlation effects; but at the expense of losing the simple qualitative description of a chemical bond that is the essence of this present manuscript.

We must keep in mind that this GVB-SOPP approximation will not always be suitable for an accurate description of a system. But for the time being we focus on systems for which it is.

As noted in Round 2, the MO method leads to delocalized orbitals that are distributed over the molecule. In the case of $CH_4$, the MOs are distributed over the 5 atoms. Linear combinations of these "canonical" MOs can be made to "localize" the orbitals into new orbitals that resemble C-H bonds, but there is no unique way in which this can be done.

For the $CH_4$ molecule, the atomic orbitals are: one 2s and three 2p orbitals. Combinations of these four atomic orbitals provide four "hybrid atomic orbitals" that point to the apices of a tetrahedron. This has led to much qualitative reasoning that has placed atomic orbitals at the center of chemical thinking. Note, however, that there is no sound theoretical justification for restricting the orbitals of a molecule to be centered on atoms. Why not allow them to move to get the lowest energy?

Most of the current calculations on molecules use large numbers of atom-centered Gaussian functions (16), instead of H atomic functions or Slater Type Orbitals (STOs) that were used in the early days of computations (17). The Gaussian orbitals are grouped into combinations that mimic atomic s, p, d, f, etc. functions, but have more variational flexibility.

Let's consider the situation for molecules containing atoms in the next row of the periodic table. The S atom is particularly interesting as it has a wide range of molecules and chemistry that has no analog in the first row of the periodic table. Since the S atom has 6 valence electrons, it would appear to be like the O atom in the first row. The analog of the water molecule, $H_2O$, is the $H_2S$ molecule. But in comparison to water, it is highly toxic. The $SF_4$ and $SF_6$ molecules are stable molecules, but $OF_4$ and $OF_6$ do not exist. The $SO_2$ molecule appears to have two S-O double bonds (18) - very different than the ozone analog, $O_3$.

This difference between S and O compounds appears to require some more fundamental understanding. Likewise, the difference between N and P and their compounds needs a better understanding. It is not that the 3s and 3p atomic orbitals of S form a chemistry analogous to what happens in the case of the 2s and 2p atomic orbitals of O.

This has led us to this series of GVB-SOPP calculations aimed at trying to understand these differences and providing a simple qualitative explanation for the differences. This started over 30 years ago (2) and was only taken up again over the last 2 years by the present authors.

The key to understanding what follows is that the number of possible covalent bonding valence orbitals is six in S and two in O. We will explain why this is so. The basic reason is that the first row of atoms, Li through Ne, has a core consisting of the two 1s orbitals that is very small in comparison to all the other atomic cores in the periodic table. For example, the S atom core consists of the 1s, 2s and 2p orbitals. Hence the first row is the exception, not the rule. Yet most of the papers and books that discuss chemistry from a theoretical point of view treat the first row as the rule and tries to explain the rest of the periodic table with those concepts.

At this point it is important to expand on some concepts that have been mentioned above. The concepts are:
1) Electron correlation effects
2) Static electron correlation effects
3) Dynamic electron correlation effects.

Rigorous MO theory (15) does not include any electron correlation effects, but additional methods (such as perturbation theory) can be added to treat electron correlation effects. However, the Heitler-London theory (12) accounts for the static electron correlation of the two electrons forming the bond.

What is static correlation and how can it be extended beyond the $H_2$ molecule? Quite simply it is the situation that obtains in a bond between two electrons, one of which is on one atom and the other electron that is on the other atom forming a Heitler-London type chemical bond. This is what the GVB-SOPP does in general for any set of covalent bonds. It is the reason we choose this approach for the calculations presented here. But what about dynamic correlation effects? The dynamic correlation effects can be treated by adding perturbation theory to the GVB-SOPP method as described in ref (19, 20).

Electron correlation effects are absolutely necessary to have a quantitative theory! The critical point for our work is to rigorously divide the static and dynamic correlation effects. Static correlation effects, as treated here, allow for a simple intuitive understanding of two-center electron bonds. Dynamic correlation effects are necessary for quantitative accuracy. A simple example of static correlation is found in the $H_2O$ molecule. There are four pairs of electrons pointing in roughly tetrahedral directions. Two arise from the two O-H bonds, the others are lone pairs pointing roughly in the other two tetrahedral directions. For each lone pair, one electron is closer to the O atom than the other -- this is in-out static correlation.

The modern valence bond theory, such as described in ref 4, for example, mix the two types of electron correlation, which make a simple interpretation of results very difficult, as illustrated in that reference.

The explanation of why the Li-Ne row is so different was offered in a paper in 1990 (2). The calculation was based on H pseudo-atoms and used a smaller basis set than used here. It started with the $SH_2$ molecule, but the H pseudo atoms could be tuned to increase the electron withdrawing capacity of these atoms, thus simulating the effect of going from H atoms to F atoms.

What was observed was that as the electron withdrawing power was increased, the S lone pairs, which were originally "in-out" correlated as in the water molecule became "angularly correlated" so that they could be bonded to other atoms. This is not possible for the water molecule because the small size of the O core cannot accommodate angularly correlated lone pairs.

# Results
## Calculational Approach

We consider calculations on the following molecules: $SF_2$, $SF_4$, $SF_6$, $SO_2$, $SO_2F_2$, $F_2SO$, $SO_3$ and $S_2O_2$. The calculations will be restricted to the GVB-SOPP approximation, which is the simplest generalization beyond MO theory that accounts for static correlation ONLY but also provides a simple explanation for chemical bonding as a generalization of the Heitler-London description of the $H_2$ molecule. We note that the calculations, mostly described in detail in the Supplementary Information are in excellent agreement with the experimental data on these molecules.

To have a meaningful comparison of the results for the various molecules considered here, it is necessary to establish a set of standards that all the computations adhere to. The standards that we establish are the following:
 a) All the calculations are for all the valence electrons of all the atoms in the molecule.
 b) The core (non-valence) electrons are treated with a newly developed "effective core potential" (8). Thus, the effective potential for O and F is for the 1s core of these atoms and the effective potential for the S atom is for the 1s2s2p core electrons.
 c) All the valence electrons are treated at the GVB-SOPP level of calculation. That is, all the electron pairs are correlated.
 d) The Gaussian basis set for all the atoms is the same in all the calculations – namely the ccpvtz type basis set developed to be optimal with the use of the ECP of (8).

This allows for a consistent and meaningful comparison among the molecules, which seems to be the first time this has been achieved.

## The $SF_2$, $SF_4$, and $SF_6$ molecules

The first GVB-SOPP calculations for the $SF_2$, $SF_4$ and $SF_6$ molecules were carried out in 1977 (3). This work treated the hyper-valency of S, which was far in advance of its time. The calculations were carried out using the experimental geometries of the molecules, necessitated by the computational capabilities of the time. The $SF_6$ molecule yielded 6 orbitals about the S atom; four were equivalent in the equatorial plane and two were different in the axial direction. In retrospect this was a warning sign, as discussed below. Had the geometry been optimized, there would be different calculated bond lengths for the equatorial and axial bonds. This would contradict the experimental results that the six bonds all have the same bond lengths. This may be an example of the failure of the GVB-SOPP approximation and a sign that resonance structures need to be added to the system wave function. The procedure for adding resonance structures has been developed (21) and applied (22) to several molecules. But we want to restrict our discussion here to GVB-SOPP cases. The real explanation for the unequal bond lengths is that the basis set used by Hay was too restrictive and when a larger basis set is used (as in the calculations presented here), the bond lengths are all equal.

To discuss bonding in a visual framework, we need to consider a few figures below that allow for a simple understanding of the bonding and the nature of the GVB-SOPP orbitals.

In Figure 1, we have a simple model of an octahedron with two schematic
orbitals along the z-axis. The $SF_6$ molecule has this geometry with six orbitals pointing along the x-, y- and z-axes.

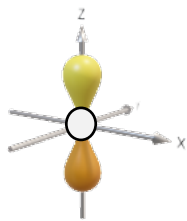

**Figure 1.** Octahedral arrangement of six S orbitals, showing only those along the z-axis – the axial orbitals. The four orbitals in the xy-plane are referred to as the equatorial orbitals.

There are two symmetric arrangements of six orbitals possible. Besides the octahedron there is also the trigonal-bipyramid arrangement. This is harder to depict, but we attempt to do so in Figure 2.

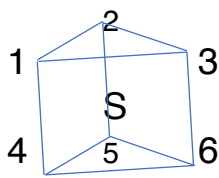

**Figure 2**. This is a 3-dimensional object (a trigonal bipyramid) that we imagine has a S atom at the center and 6 apices to which orbitals point from the center S atom.

The $SF_2$ molecule according to Hay has two S-F single bonds, along the positive x and y axes of Figure 1. Alternatively, the orbitals could be as shown in Figure 2, where we will say the F atoms are at the apices 3 and 6. There are two orbitals from S that point to apices 1 and 4, which are described by a covalent bond as in the hydrogen molecule. Likewise, there are two orbitals that point from the S atom to the apices 2 and 5 and are coupled as in the hydrogen molecule, these are the two S lone pairs. This is a very different perspective than is offered by MO theory or DFT. How can we explain this difference?

Over the last 20-30 years, DFT has become the most popular approach to computational calculations for molecular systems. As in MO theory, there are no unique orbitals that help in understanding chemistry from a simple chemical valence bond perspective. This is a critical point, although DFT and MO theories may provide some unique perspectives, they cannot provide a valence bond description! Our approach is designed to provide that description and thus complement other approaches which cannot provide that perspective. Further, our approach provides an excellent starting point for a more quantitative description of molecules as discussed in the Conclusions section of the paper.

However, MO and DFT calculations do yield $SF_6$ bonds that are equivalent and in reasonable agreement with the experimental S-F bond length, especially for the DFT, which was calculated (here) to give a bond length of 1.589 Å (see Table 2 below).

One of the unsatisfying aspects of the DFT approach is that there is no unique DFT exchange-correlation potential. There are dozens or more such potentials that are used according to the preferences of the authors of calculations on specific systems.

Let's return to our discussion of the $SF_2$, $SF_4$ and $SF_6$ molecules. All the calculations presented in this paper were carried out with the Jaguar software of Schrödinger, Inc. (9). The geometry of the $SF_2$ molecule is a pseudo-octahedron as in Figure 1, where the two S-F bonds are along the positive x- and y- axes at the locations marked 1 and 2 in Figure 3. There are two angularly correlated orbitals that are paired as in the Heitler-London treatment of the hydrogen molecule; one along the z-axis (positions 5 and 6 in Figure 3) and the other at the positions 3 and 4 in Figure 3.

For the $SF_4$ molecule, the addition of 2 F atoms to the $SF_2$ molecule is such that the new bonds are at positions 5 and 6 along the z-axis, one above and one below the equatorial plane (see, Figure 3). Then the $SF_6$ molecule is formed by bonding two F atoms to the angularly correlated orbitals in the equatorial plane (positions 3 and 4 in Figure 3). The experimental geometries are given in Table 1. The calculated optimal geometries are given in Table 2.

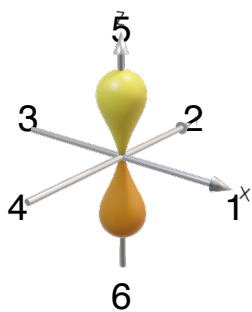

**Figure 3.** The octahedral geometry of the 6 valence orbitals of S. The axial orbitals (labelled 5 and 6) are schematically shown, the others are implied. $SF_6$ has all the positions occupied by F atoms. The $SF_4$ molecule has positions 1, 2, 5 and 6 occupied by F atoms and positions 3 and 4 are an angularly correlated pair. $SF_2$ has positions 1 and 2 occupied by F atoms and 5 and 6 are now also an angularly correlated pair. Atom 7 is S, but not shown for clarity. In all the Tables below, the numbering scheme in the Figures 1 to 3 are used to specify the bonds and the bond angles.

**Table 1.** The $SF_2$, $SF_4$ and $SF_6$ calculated total Energies from the GVB-SOPP calculations at the experimental geometries are as follows:

| Molecule | Bond Length | Bond Angles | Total Energy (ha) |
|---|---|---|---|
| $SF_2$ | 1.59 Å (7-1 & 7-2) | 98.05° (1-7-2) | -58.09920 |
| $SF_4$ | 1.65 Å (7-5 & 7-6) | 173.1° (5-7-6) | -106.19811 |
|  | 1.55 Å (7-2 & 7-3) | 87.8° (1-7-2) |  |
| $SF_6$ | 1.56 Å (all) | 90° (1-7-5, etc.) | -154.32770 |

$SF_6$ experimental geometry, see Wikipedia
$SF_2$ and $SF_4$ experimental geometries see ref. 18

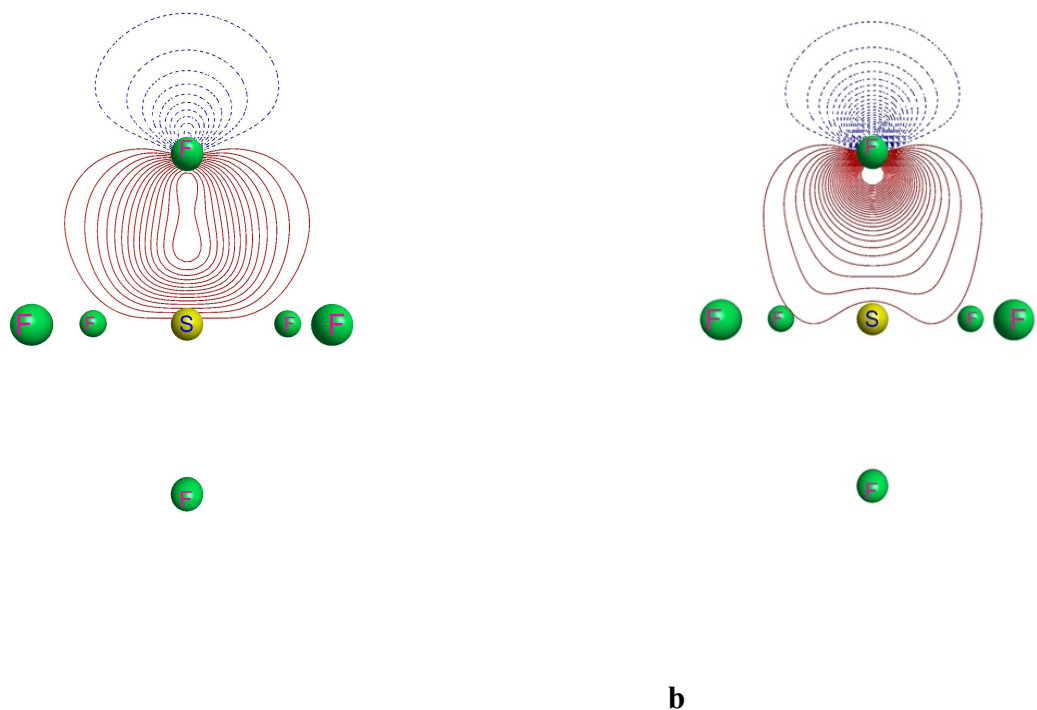

        a                                                       b

**Table 2.** The $SF_2$, $SF_4$ and $SF_6$ GVB-SOPP optimized Geometries and Total Energies are as follows:

| Molecule | Bond Lengths | Bond Angles | Total Energy (ha) |
|---|---|---|---|
| $SF_2$ | 1.577 Å (7-1 & 7-2) | 97.56° (1-7-2) | -58.09938 |
| $SF_4$ | 1.619 Å (7-5 & 7-6) | 171.66° (5-7-6) | -106.21036 |
|  | 1.521 Å (7-2 & 7-3) | 87.44° (1-7-2) |  |
| $SF_6$ | 1.542 Å | 90° (1-7-5, etc.) | -154.32944 |
| "   (DFT) | 1.589 Å | " | NA |

It is very important to establish how the GVB-SOPP calculations confirm our anticipated descriptions of the orbital bonding in the molecules treated here. This is best done by showing the calculated orbitals forming the bonds. We will avoid doing this for all the cases, as shown in

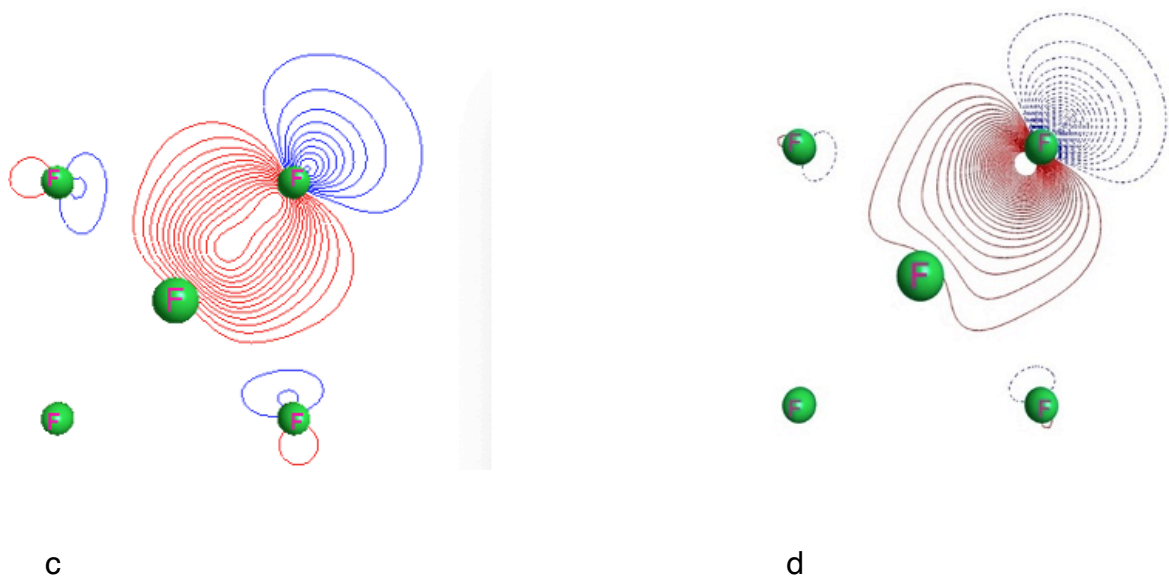

c                            d

**Figure 4.** The GVB-SOPP orbitals for the $SF_6$ molecule. One of the two equivalent axial S-F bonds are shown in panels a, and b. One of the 4 equivalent equatorial S-F bonds in the x-y plane are shown in c and d.

them for a few will establish the point. We start with the situation for the $SF_6$ molecule, where we say that there are six equivalent bonds about the central S atom.

In Figure 4, we provide four orbital plots which show that there are 6 equivalent S-F bonds in the $SF_6$ molecule. Note that a bond in the GVB-SOPP method consists of two orbitals mainly centered on separate atoms that form the bond. In this case, one is mainly centered on the S atom (as in Figure 4a) and the other mainly centered on the F atom (as in Figure 4b). This is the axial bond along the positive z-axis.

The $SF_4$ molecule is next in our discussion. It can be thought of as being a result of removing the two F atoms in the axial positions (see upper half of Figure 4 for one of these). This creates two lone pair orbitals on S that are angularly correlated (shown in Figure 5 a and b). There remain four S-F bonds. One of two equivalent S-F bonds is shown in panels c and d. The other two S-F bonds are at roughly 90° (orbitals not shown). These orbitals are slightly different than those in panels c and d, due to the position of the S lone pair.

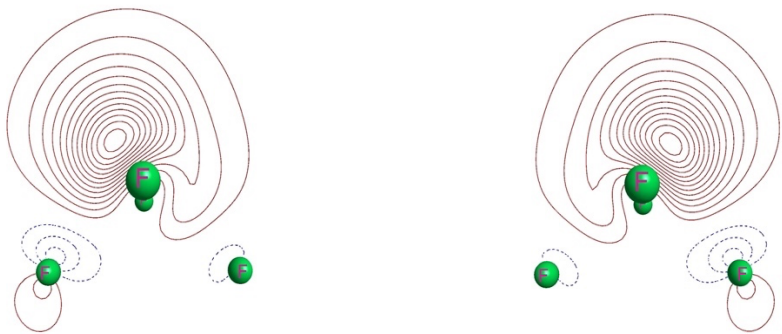

a             b

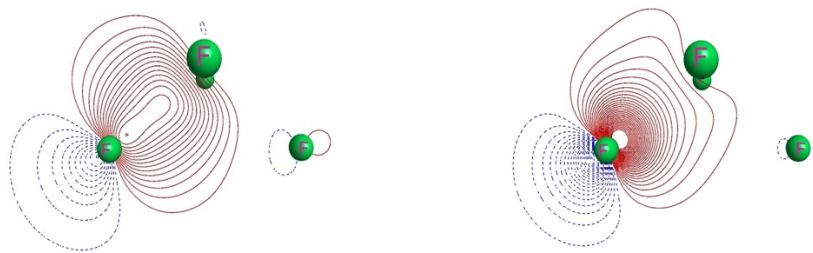

c             d

**Figure 5.** The GVB-SOPP orbitals for the $SF_4$ molecule. The angularly correlated lone pair is shown in panels a and b. One of the two equivalent axial S-F bonds in this plane are shown in panels c and d.

The last molecule to discuss here is the $SF_2$ molecule. It consists of two equivalent S-F bonds and two equivalent S lone pairs accounting for the six valence orbitals about the S atom. The lone pairs on S are like the one shown in Figure 5 for $SF_4$. The orbitals about the S atom rearrange slightly from the octahedral arrangement of Figure 1 into the trigonal bipyramid configuration of Figure 2. We show the orbitals for one of the S-F bonds in Figure 6, panels a and b. This completes the discussion of the $SF_n$ molecules. Note the transferability of the GVB-SOPP orbitals (shown in the Figures above) among the $SF_6$, $SF_4$ and $SF_2$ molecules.

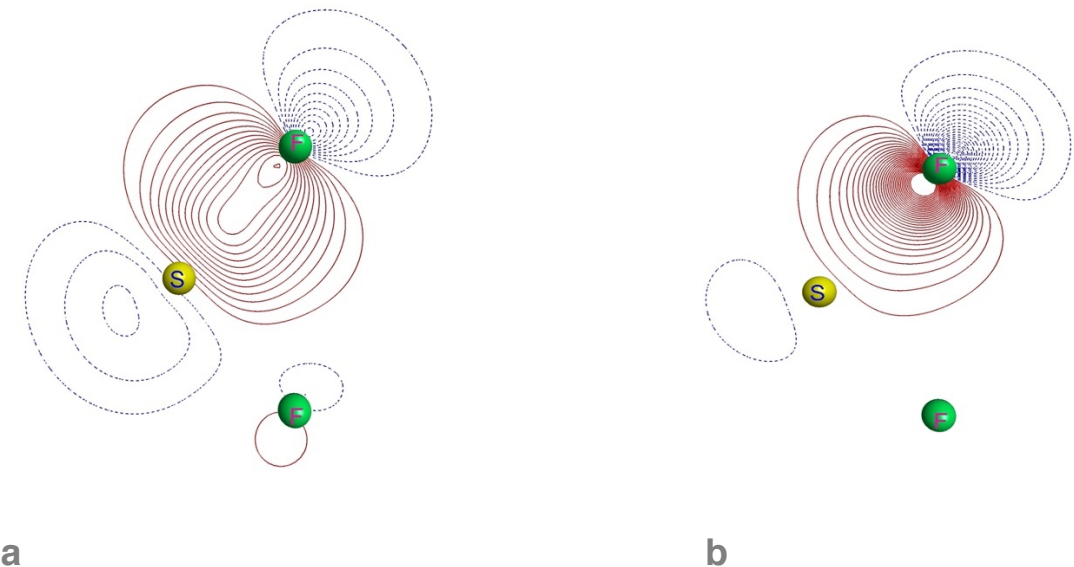

**Figure 6.** The GVB-SOPP orbitals for the $SF_2$ molecule. Orbitals of one of the S-F bonds are shown in panels a and b.

There are several other molecules for which we have carried out the same type of quantitative calculations and they confirm the general findings found for the molecules discussed above. We discuss those other molecules, $SO_2$, $SO_2F_2$, $F_2SO$, $SO_3$ and $S_2O_2$ below.

## Results for the $SO_2$ molecule

The $SO_2$ molecule has the anticipated orbital structure displayed in Figure 2. One of the O atoms forms a double bond with the S orbitals pointing to 3 and 6, and the other O atom forms a double bond with the S orbitals pointing to 1 and 4. The remaining two orbitals of S, pointing to 2 and 5 are paired angularly correlated orbitals. The S-O bonds are each 1.432 Å. See Tables 3 and 4.

**Table 3.** $SO_2$ GVB-SOPP calculation at experimental geometry

| Molecule | Bond Lengths | Bond Angle (O-S-O) | Total Energy (ha) |
|---|---|---|---|
| $SO_2$ | 1.432 Å | 119.5° | - 41.637920 |

**Table 4.** $SO_2$ GVB-SOPP calculation at optimized geometry

| Molecule | Bond Lengths | Bond Angles (O-S-O) | Total Energy (ha) |
|---|---|---|---|
| $SO_2$ | 1.407 Å | 118.4° | - 41.639939 |

Some of the unique GVB-SOPP orbitals of the $SO_2$ molecule are shown in Figure 7. Our expectations of the bonding are verified by the actual calculations. The S lone pairs and the S-to-O bonds appear to be highly transferable - both in terms of bond lengths and the orbitals that describe them.

Note that we encounter "bent bonds" in Figure 7 (a and b) for the first time in our discussions here. Such bonds were part of the chemical thinking in the first few decades of the 20$^{th}$ Century but were supplanted by sigma/pi double bonds based on MO theory thinking. They were taken up again (2) when it was found that the energy of the "bent-bonds" description was lower in energy than the sigma/pi description. This has been confirmed by many others since then. In the calculations presented here, bent bonds are always found to be lower in energy for the molecules discussed here. Either bent bonds or sigma/pi retain the full symmetry of the molecule.

An interesting possibility for the reactivity of an excited state of $SO_2$ and its chemistry has been suggested by the recent work of Anglada et al (23). Their suggestion is that the angularly correlated lone pair (in our terminology) can be unpaired to provide two reactive sites for forming new chemical bonds.

This triplet state entity could possibly be important in chemical reactions such as those discussed in that paper (23). It may be necessary, however, to expand our thinking beyond GVB-SOPP, in order to get an accurate description of new bond formation (breaking an old bond and forming a new bond). This is a subject we will take up in a future paper. However, for the present we give some GVB-SOPP results for the excited triplet state of the $SO_2$ molecule.

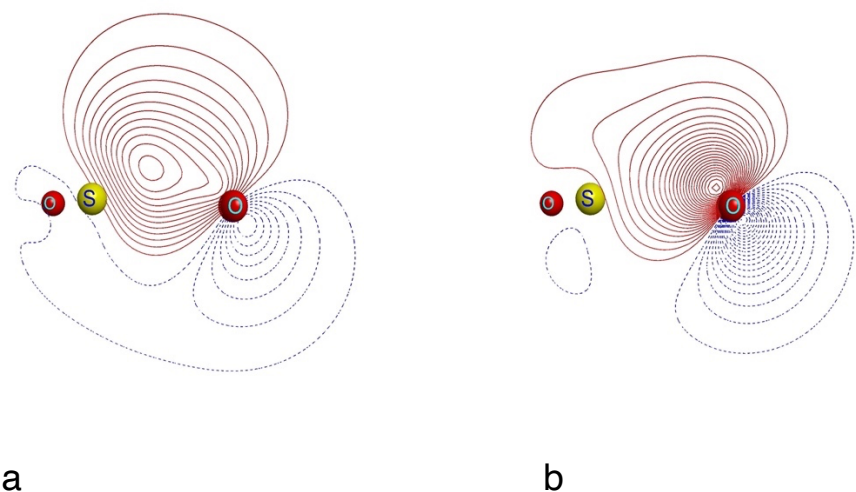

a                              b

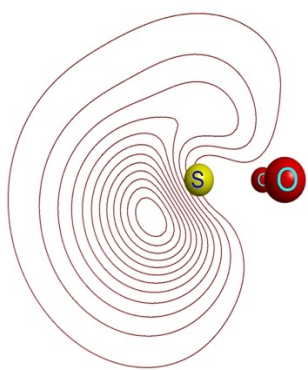
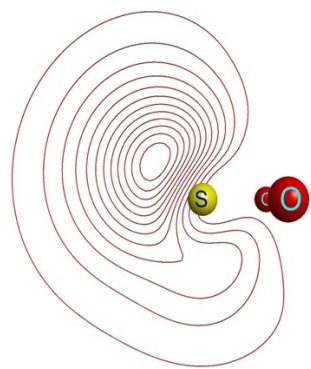

c                                                    d

**Figure 7.** The GVB-SOPP orbitals of the $SO_2$ molecule. Panels a and b show one component of an S-O double bond, the other component is a mirror image reflection about the $SO_2$ plane. The S angularly correlated lone pair is shown in panels c and d.

We show, in Figure 8, the orbitals of the excited triplet state. The two S-O double bonds are essentially the same as in the ground state. In Figure 8 there are two equivalent descriptions of the orbitals of the triplet state shown. The first orbitals, in panels a and b, are the sigma, pi description of the triplet state. A linear combination of these orbitals, which does not change the energy of the molecule, are shown in panels c and d of Figure 8. The latter orbitals look very similar to the original angularly correlated lone pair orbitals (see Figure 7 c and d) of the $SO_2$ molecular ground state.

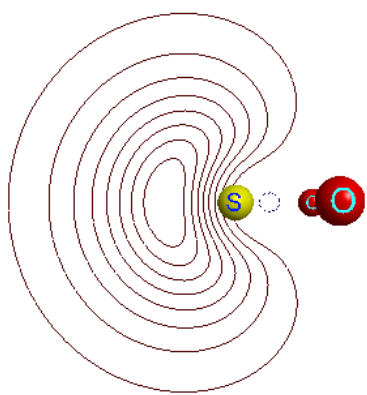
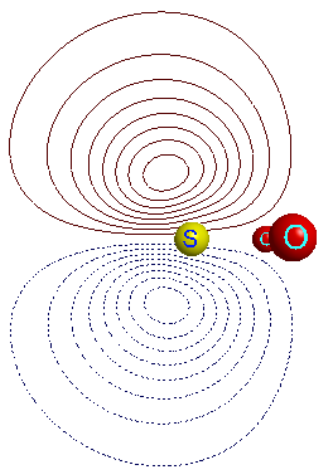

a                                                    b

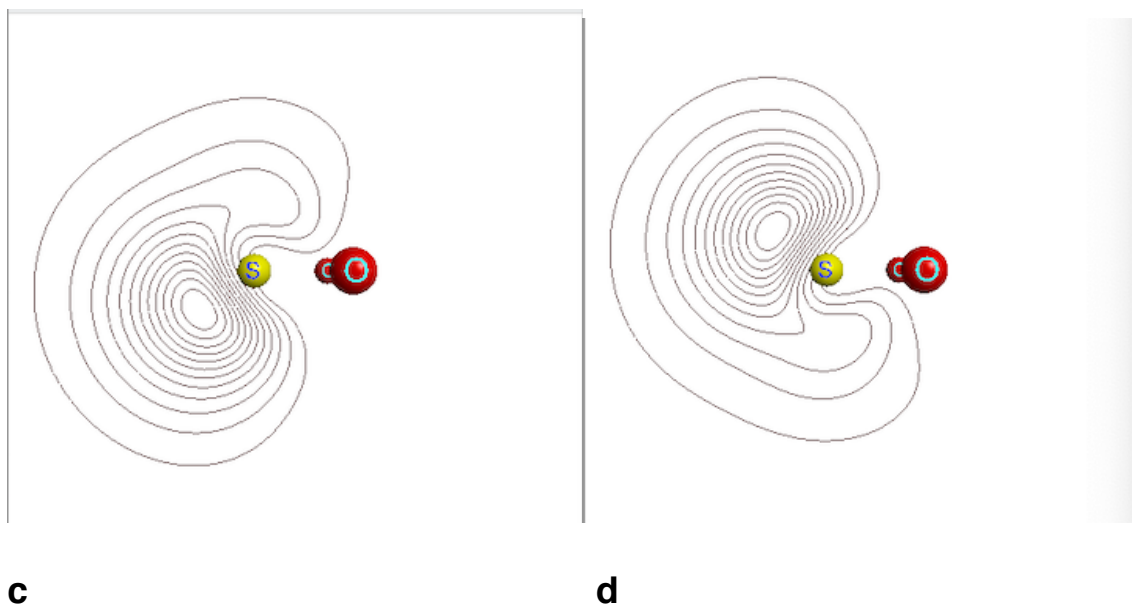

**Figure 8.** The two versions of the triplet state orbitals of the SO$_2$ molecule. Panels a and b are the sigma, pi version of the orbitals and Panels c and d are a linear combination of those orbitals that look very similar to the original angularly correlated lone pair orbitals of the ground state. Both versions yield identical energies.

## SO$_2$F$_2$ and SOF$_2$ molecules

Building on the description of the SO$_2$ molecule, we expect the SO$_2$F$_2$ molecule to have the two angularly correlated orbitals of SO$_2$ (see Figure 7 c and d) forming two S-F single bonds. The S-O bonds are each 1.405 Å and the S to F bonds are each 1.53 Å (see Tables 5 and 6).

**Table 5.** SO$_2$F$_2$ GVB-SOPP calculation at experimental geometry

| Molecule | Bond Lengths | Bond Angles | Total Energy (ha) |
|---|---|---|---|
| SO$_2$F$_2$ | 1.530 Å (S-F) | 123.97° (O-S-O) | -89.763148 |
|  | 1.405 Å (S-O) | 96.12° (F-S-F) |  |

**Table 6.** SO$_2$F$_2$ GVB-SOPP calculation at optimized geometry

| Molecule | Bond Lengths | Bond Angles | Total Energy (ha) |
|---|---|---|---|
| SO$_2$F$_2$ | 1.509 Å (S-F) | 124.14° (O-S-O) | -89.765717 |
|  | 1.387 Å (S-O) | 95.46° (F-S-F) |  |

Note how the fact that there are 6 valence orbitals of S, provides for a simple explanation of the structure of this molecule. This contrasts to the atomic orbital description based on concepts from the first row of the periodic table.

The $SOF_2$ molecule is expected to have an S-O double bond and two S-F single bonds. The remaining two S orbitals are expected to form an angularly correlated pair. The orbitals around the S atom are expected to be in a trigonal prism geometry as shown in Figure 2. This is what we expect based on arguments made above, and it is what we find from the detailed quantum calculations. The experimental S-F bond length is 1.59 Å compared to that of 1.59 Å for the $SF_2$ molecule. The S-O bond length is 1.41 Å as compared to the S-O bond length of 1.43 Å in the $SO_2$ molecule. See Tables 7 and 8.

**Table 7.** $SOF_2$ GVB-SOPP calculation at experimental geometry

| Molecule | Bond Lengths | Bond Angles | Total Energy (ha) |
|---|---|---|---|
| $SOF_2$ | 1.585 Å (S-F) | 106.82° (O-S-F) | -73.94924 |
| | 1.413 Å (S-O) | 92.83° (F-S-F) | |

**Table 8.** $SOF_2$ GVB-SOPP calculation at optimized geometry

| Molecule | Bond Lengths | Bond Angles | Total Energy (ha) |
|---|---|---|---|
| $SOF_2$ | 1.551 Å (S-F) | 106.50° (O-S-F) | -73.95179 |
| | 1.402 Å (S-O) | 92.51° (F-S-F) | |

The calculations confirm our expectations and there seems to be no need to show orbitals that would be redundant with those already displayed for other molecules.

## Results for $SO_3$, and $S_2O_2$ molecules

The $SO_3$ molecule is expected to be planar and have 3 S-O double bonds, with the S orbitals arranged in a trigonal prism arrangement as in Figure 2. The experimental S-O bonds in this molecule are all equal and are 1.42 Å. Hence, the $SO_2$, $SO_3$ and $SOF_2$ molecules have S-O bond lengths of 1.43 Å, 1.42 Å and 1.41 Å. See Tables 9 and 10 for $SO_3$ results.

**Table 9.** $SO_3$ GVB-SOPP calculation at experimental geometry

| Molecule | Bond Lengths | Bond Angles | Total Energy (ha) |
|---|---|---|---|
| $SO_3$ | 1.418 Å (S-O) | 120.0° (O-S-O) | -57.42393 |

**Table 10.** $SO_3$ GVB-SOPP calculation at optimized geometry

| Molecule | Bond Lengths | Bond Angles | Total Energy (ha) |
|---|---|---|---|
| $SO_3$ | 1.397 Å (S-O) | 120.0° (O-S-O) | -57.42698 |

The $S_2O_2$ molecule, the SO dimer, is of considerable interest because of its physical importance in the atmospheric chemistry of the planet Venus (24-27). To visualize the bonding in this molecule, it requires expanding our view of the bonding beyond that in Figures 1 and 2. Let's start with Figure 2 with the 6 S orbitals pointing to the labelled vertices. To consider two S atoms, imagine bringing another trigonal prism, as in Figure 2 and place that trigonal prism to the right of the original trigonal prism. Then a S-S double bond is formed by having the S orbitals at vertices 1 and 4 of the right trigonal prism bond to the S orbitals at vertices 3 and 6 of the left trigonal prism. If we allow an O atom to make a double bond with the S on the left, using S orbitals 1 and 4, then there are two options for making a S to O double bond with the S atom to the right. One option (the cis configuration) is to bond to the orbitals 3 and 6, the other option (the trans configuration) is to bond to the orbitals 2 and 5 on the right S atom. There are two angularly correlated pairs of electrons, one on each sulfur atom.

There does not appear to be any experimental data on the geometries of the cis and trans structures of these molecules. However, there are many previous calculations on these molecules, suggesting that the cis conformer is lower in energy than the trans conformer. Although a complete description of the bonding in these molecules is still lacking, we use the knowledge that we have gained thus far on other molecules presented here to suggest the bonding in these two isomers of the SO dimer.

The lowest energy geometry is in the cis configuration and the O-S-S bond angle is 113º. The reason that the cis configuration is more stable appears to be that there are more unfavorable lone pair-lone pair interactions in the trans configuration than in the cis configuration.

**Table 11**. SO cis-dimer GVB-SOPP calculation at optimized geometry

| Molecule | Bond Lengths | Bond Angles | Total Energy (ha) |
| --- | --- | --- | --- |
| OSSO (cis) | 1.439 Å (S-O) | 116.8° (O-S-S) | -51.569415 |
| | 1.915 Å (S-S) | | |

**Table 12**. SO trans-dimer GVB-SOPP calculation at optimized geometry

| Molecule | Bond Lengths | Bond Angles | Total Energy (ha) |
| --- | --- | --- | --- |
| OSSO (trans) | 1.446 Å (S-O) | 110.30° (O-S-S) | -51.559672 |
| | 1.930 Å (S-S) | 179.48° (O-S-S-O) | |

A summary of all the results presented in this paper is given in Table 13 below.

**Table 13.** Summary of the GVB-SOPP calculations presented in this paper.

| Bond Angles | Index | Molecule | SS Bond | SO Bond | SF Bond | S lp | O lp | F lp | Tot E (expt) | Tot E (opt) |
|---|---|---|---|---|---|---|---|---|---|---|
| | 1 | **SF$_6$** | 0 | 0 | 6 | 0 | 0 | 18 | -154.32770 | -154.32944 |
| 5-7-6: 180° | expt | " | | | 1.564 Å | | | | | |
| 5-7-6: 180° | opt | " | | | 1.542 Å | | | | | |
| | 2 | **SF$_4$** | 0 | 0 | 4 | 1 | 0 | 12 | -106.20095 | -106.21036 |
| 1-7-2: 87.8° | expt | " | | | 1.545 Å | | | | | |
| 5-7-6: 173.1° | expt | " | | | 1.646 Å | | | | | |
| 1-7-2: 87.44° | opt | " | | | 1.518 Å | | | | | |
| 5-7-6: 171.66° | opt | " | | | 1.619 Å | | | | | |
| | 3 | **SF$_2$** | 0 | 0 | 2 | 2 | 0 | 6 | -58.09909 | -58.09938 |
| 1-7-2: 98.3° | expt | " | | | 1.589 Å | | | | | |
| 1-7-2: 97.6° | opt | " | | | 1.577 Å | | | | | |
| | 4 | **SO$_2$** | 0 | 4 | 0 | 1 | 4 | 0 | -41.63792 | -41.63994 |
| O-S-O: 119.5° | expt | " | | 1.43 Å | | | | | | |
| O-S-O: 118.4° | opt | " | 0 | 1.41 Å | | | | | | |
| | 5 | **SO$_2$F$_2$** | 0 | 4 | 2 | 0 | 4 | 6 | -89.763 | -89.765 |
| O-S-O: 124.0° | expt | " | | 1.405 Å | 1.53 Å | | | | | |
| F-S-F: 96.1° | expt | " | | | | | | | | |
| O-S-O: 124.1° | opt | " | | 1.40 Å | 1.55 Å | | | | | |
| F-S-F: 95.4° | opt | " | | | | | | | | |
| | 6 | **SO$_3$** | 0 | 6 | 0 | 0 | 6 | 0 | -57.41516 | -57.42699 |
| O-S-O: 120° | expt | " | | 1.418 Å | | | | | | |
| O-S-O: 120° | opt | " | | 1.397 Å | | | | | | |
| | 7 | **F$_2$SO** | 0 | 2 | 2 | 1 | 2 | 6 | -73.949 | -73.952 |
| F-S-O: 107.1° | expt | " | | 1.405 Å | 1.53 Å | | | | | |
| F-S-F: 92.6° | expt | " | | | | | | | | |
| F-S-O: 106.5° | opt | " | | 1.402 Å | 1.55 Å | | | | | |
| F-S-F: 92.5° | opt | " | | | | | | | | |
| O-S-S 116.8° | 8 | **S$_2$O$_2$ -c** | 2 | 4 | 0 | 2 | 4 | 0 | ---- | -51.5694 |
| | | " | 1.915 Å | 1.439 Å | | | | | | |
| | 9 | **S$_2$O$_2$ -t** | 2 | 4 | 0 | 2 | 4 | 0 | ---- | -51.5596 |
| O-S-S 110.3° | | | 1.930 Å | 1.446 Å | | | | | | |
| O-S-S-O 179.48° | | | | | | | | | | |

## Discussion

It's important to compare the results presented above for the $SO_2$ molecule with some of the traditional reasoning on the bonding in the $SO_2$ molecule. An example paper (4) makes several assumptions that from the perspective of our discussion above might be questionable. However, these assumptions are by no means unique to this paper (4).

1) It assumes that the bonding can be explained using reasoning from the Li-Ne row of the periodic table.
2) It restricts the discussion of bonding to sigma/pi orbitals although it has been shown that a lower energy is obtained by using bent bonds ($\Omega$ - bonds).
3) Although it uses GVB calculations, it does not use the GVB-SOPP approximation and therefore does not allow a simple valence bond pairing description (see our discussion above on the bond pairing in the $CH_4$ molecule).

This leads to a discussion that finds the observed differences in ozone ($O_3$) as compared to $SO_2$ difficult to explain (4). We maintain that this difference is dictated by the size of the atomic cores as noted in the Abstract and discussed in the Introduction. Only 4 orbital pairs (two bonds and two radially correlated lone pairs) can be accommodated in the valence shell of the atoms Li-Ne. This contrasts with the Na-Ar row which can accommodate 6 (or more) orbitals because of its much larger core size. Trying to explain S bonding in terms of O bonding is indeed a challenging task. This was also noted for the difference between P and N bonding in an article by Messmer in 1990 that is discussed later.

This situation arose because of the historical development of quantum mechanics and its focus on the chemistry of bonding associated with the elements H through Ne. Organic chemistry and biochemistry were quite a challenge in themselves and taking on the rest of the periodic table was clearly for the faint-hearted, with some notable exceptions.

The delay in addressing other rows in the periodic table from a more rigorous perspective was due to the complexity of the quantum calculations necessary. When there were more than four bonds (orbital pairs) in a molecule, as in $SF_6$, these molecules were described as "hypervalent," meaning they contained more than the expected four bonds. A general treatment of hypervalent molecules is still lacking to this day.

The present work is a step toward opening that discussion and providing an approach which needs to be tested with many more calculations than presented here. However, this paper provides a framework that can be rigorously tested to see if the concepts presented hold up over time. This is, after all, the nature of the scientific method.

## Conclusions

We submit that the calculations presented in this paper provide strong evidence in support of local bonds between pairs of atoms. The computed orbitals (and their orbital plots) from the molecules presented here strongly suggest that all S to O double bonds look very much the same and have bond lengths that are very much the same; the S to F bonds all have orbitals that describe them that are very much the same and have bond lengths that are very much the same.

The size of the atomic cores in the atoms Li-Ne has an order of magnitude smaller volume than those of the next row in the periodic table and presumably in the rest of the periodic table (perhaps to an even greater extent). This is what allows (in the rest of the periodic table) for more covalent bonds to be formed than is allowed by the "Octet Rule."

As noted above, a general treatment of hypervalent molecules is still lacking to this day. To illustrate this point we refer to a Wikipedia article (29). Note: in this article there is no simple set of unifying concepts presented. There are many viewpoints given with the only common theme being that hypervalent molecules are different than those described by the Octet Rule.

How is it possible, after all the advances in physics, chemistry, and computations over the last several decades, that this is the current situation?

Why, after so many years, are we still trying to use those concepts that worked for Li-Ne to describe the rest of the periodic table?

Perhaps a fixation on past successes of describing Li-Ne chemistry can be the reason! Understandable. Or is it that some of the "really fundamental" concepts have been forgotten or mis-interpreted? Forgetting that the only accurate solution of the Schrödinger Equation is for the H atom and that so many of the concepts currently used are derived from the H atom solution, not from molecular solutions to the Schrödinger equation. This was understood decades ago – but forgotten. The Heitler-London description of the $H_2$ molecule is the only accurate solution of a molecule – but as pointed out above, it has some approximations – BUT it takes account of the antisymmetric principle, which is crucial.

Another interesting question is how the name of Linus Pauling has almost been forgotten in terms of his contributions to Valence Bond Theory (VBT). Recall that he initiated a long series of articles entitled "The Nature of the Chemical Bond" beginning in 1931 (30). These articles established VBT as the foremost means of understanding bonding in molecules and solids. This was followed by his book in 1948 (31), which to many was the bible on the chemical bond. How could all of this seminal work be forgotten. One explanation is that it occurred before the advent of modern computers and computations.

Pauling was aware that the overlap of atom-based orbitals could not account for the changes in electronic re-distribution taking place in bond formation. To account for this change in charge distribution, he chose to invoke "resonance" structures to accomplish the task. Unfortunately, the number of these resonance structures soon became extremely large and there were no computations to guide choices. Molecular Orbital Theory was able to account for this redistribution via SCF calculations. As noted above VB theory is much more difficult to implement and required some significant *ad hoc* approximations.

Thus, the issue of resonance structures was considered by many in the computational community as a final reason to abandon VBT in favor of MOT. It was not until the generalization of the Heitler-London approach was discovered in 1972 (5), that it became abundantly clear that these many resonance structures of Pauling were automatically incorporated into the SCF procedure of the GVB-SOPP method.

What have we learned from the results presented here? In brief, the learnings are:
1. controlled approximations in a specific order are necessary
2. e.g., do not mix static and dynamic correlation effects in the same calculations; keep them separate
3. as the GVB-SOPP approximation has ONLY static correlation and a simple Valence Bond interpretation (a covalent bond between two atoms), do this first; it also describes breaking and forming covalent bonds; this is the approach we have presented in this paper
4. there are some chemical reactions that require a more general GVB solution, but such generalizations should be restricted to the simplest such approach; this will be described in a future publication.
5. after all these factors have been accounted for, add the dynamic correlation effects that will be necessary to get accurate energies and reactive potential surfaces. This has previously been done and the methodology published (19, 20).

We believe this work has substantially improved our ability to have a simple bonding interpretation (covalent bonds between pairs of atoms) and provide a starting point for more accurate methods for including dynamic correlation effects. We believe this is the first approach to provide a roadmap for describing "hypervalent molecules."

**Author Information**
  **Corresponding Authors**

**Robert B. Murphy** – Schrödinger, Inc., 120 West 45th Street, New York,     New York 10036, United States; orchid.org/0000-0001-5967-5932; Email: rob.murphy@schrodinger.com

**Richard P. Messmer** – RPM Global Associates, 641 Grooms Road, Suite 106, Clifton Park, New York 12065, United States; orchid.org/ 0000-0001-9654-1159; Email: rpm_ga@yahoo.com

**Notes**
The authors declare no competing financial interest.

**Acknowledgments**
  We are extremely grateful to Prof. William A. Goddard III of Caltech for his creation of the GVB methods and particularly the GVB-SOPP method which have been an inspiration to us for decades.
  We wish to express our gratitude to Dr. Yixiang Cao of Schrödinger, Inc. for programming the new Effective Core Potentials into the standard Jaguar suite of quantum codes for general use.